\documentstyle[12pt]{article}
\begin{document}

\begin{titlepage}
\centerline{\normalsize January 1997\hfill AS-ITP- 97-01,  DOE/ER/01545}
\vspace{1.0cm}
\begin{center} \begin{Large} \begin{bf} 
Nonfactorizable Contributions to 
Hadronic Matrix Elements by Use of Twist Wave Functions
\end{bf} \end{Large} \end{center}
\vspace{1cm}
\begin{center}
    W.\ F.\ Palmer$^a$ and Y.\ L.\  Wu$^{a,b}$  \\
      \vspace{0.3cm}
        $^a$Department of Physics, The Ohio State University
\footnote{Supported in part by the US
            Department of Energy under contract DOE/ER/01545.}, \\
        Columbus, Ohio 43210, USA\\
 \vspace{0.3cm}
        $^b$Institute of Theoretical Physics \\
       Chinese Academy of Sciences \\
      Beijing, 100080, P.R. China \\
        \end{center}
  \vspace{1cm}
\noindent {\bf Abstract}\\
\\
A general approach for evaluating the nonfactorizable contributions to 
the hadronic matrix elements is described in terms of the higher twist wave
functions of the mesons. An example is illustrated for the 
$B^{0}\rightarrow \pi^{+}\pi^{-}$ decays.

\parbox[t]{\textwidth}{ }
\end{titlepage}

\newpage

  The factorization approximation has conventionaly been 
used to describe the two-body
nonleptonic decays of D and B mesons.   
However, it has been observed \cite{BSW} that the naive factorization 
approach to nonleptonic heavy meson decays is incompatible with 
experimental data. The available data for D-meson decays seems to 
support the so-called rule of discarding $1/N_{c}$ corrections. 
This rule  has been formulated in ref. \cite{BGR}, which is in contrast
to the standard prescription that keeps the $1/N_{c}$ term.  The 
dynamical explanation of this approximate cancellation of $1/N_{c}$ terms 
in the D decays has been presented in ref. \cite{BS}.  From such an 
explanation, it appears that discarding $1/N_{c}$ terms may not 
necessarily be a universal rule. The recent experimental data on B-meson 
decays suggests that discarding $1/N_{c}$ terms is not a universal rule
for all channels. To understand these problems, one needs to see what  
approximation is being made in the naive factorization approach and how to 
systematically improve upon the approach. It will be seen that the issues 
concern the nonfactorizable contributions of the operator of 
two color-octet currents.

 The leading nonfactorizable contributions to hadronic matrix element
arise from the soft gluon interactions. It turns out, such 
soft gluon contributions to  hadronic weak decays of mesons 
can be further factorized in terms of twist wave functions (TWF) of 
the mesons. We shall describe in this note a general approach for their  
evaluation.  

As an illustration, let us consider an effective QCD-corrected 
weak Hamiltonian with two familiar operators $O_{1}$ and $O_{2}$. 

\begin{equation}
H_{eff} \sim c_{1}(\mu) O_{1} + c_{2}(\mu) O_{2}
\end{equation}  
where $c_{i}(\mu)$ are the Wilson coefficient functions and $O_{i}$ the 
effective four quark operators.  For $B\rightarrow \pi \pi$ decays, 
$O_{i}$ are given by    
\begin{eqnarray} 
O_{1} & = & (\bar{u}b)_{V-A} (\bar{d}u)_{V-A}, \nonumber \\ 
O_{2} & = & (\bar{u}_{\alpha}b_{\beta})_{V-A} 
(\bar{d}_{\beta}u_{\alpha})_{V-A} 
\end{eqnarray}
In the factorization approximation, the hadronic matrix element
is factorized into the product of two matrix elements of singlet 
currents, which are evaluated in terms of decay constants and formfactors, 
and characterized by the parameters $a_{1}^{(o)}$ and $a_{2}^{(o)}$. 
These parameters are related to the Wilson coefficients $c_{i}$

\begin{equation}
a_{1}^{(o)} = c_{1} + \frac{1}{N_{c}} c_{2}, \qquad 
a_{2}^{(o)} = c_{2} + \frac{1}{N_{c}} c_{1}.
\end{equation}
 
     This naive factorization fails to describe the
nonleptonic charm meson decays.   The approximation 
neglects the contributions from the product of two color-octet 
currents, 

\begin{equation}
O_{1}^{c}  =  2(\bar{d}t^{a}b)_{V-A} (\bar{u}t^{a}u)_{V-A}, \qquad  
O_{2}^{c} = (\bar{u}t^{a}b)_{V-A} (\bar{d}t^{a}u)_{V-A} 
\end{equation}
Where $t^{a}$ ($a=1,\cdots, 8$) are
SU(3)$_{c}$ generators, and the color indices in the operators $O^{c}_{i}$ 
have be omitted. It becomes manifest that in the naive 
factorization approach at this 
level the contribution to the $B\rightarrow \pi^+ \pi^-$ decay
from the operators $O_{2}^{c}$ vanishes due to the mismatching of color.  

 Using the algebraic relation $t^{a}_{\alpha \beta} t^{a}_{\beta'
\alpha'} = \frac{1}{2} (\delta_{\alpha \alpha'} \delta_{\beta \beta'} -
\frac{1}{N_{c}} \delta_{\alpha \beta} \delta_{\alpha'\beta'})$, and applying 
the Fiertz transformation, one easily obtains the following identities
\begin{equation} 
O_{i} \equiv   \frac{1}{N_{c}} O_{j} + O_{i}^{c}
\end{equation}
where (i,j) represents the pair (1,2) and
$N_{c}$ is the number of colors.  The leading nonzero contribution of the
operator $O_{2}^{c}$ to the $B\rightarrow \pi^+ \pi^-$ decay
arises from the soft gluon emission which recovers the color matching of
the octet currents to the hadron. The resulting new effective
operators due to the soft gluon  interaction can be written as a
product of two color-singlet currents again. The leading nonzero
contribution is given by  

\begin{equation}  
A_{O_{2}^{c}}= -i\int dx < \pi^{+}(p_{+}) \pi^{-}(p_{-}) | T\ (j(x) 
O_{2}^{c}(0) ) | B(p) > 
\end{equation}
where $j(x) = \bar{d}(x) A(x) d(x)$ with $A(x)\equiv 
g_{s} A(x)^{a}_{\mu} t^{a} \gamma^{\mu}$ (other quark flavors contribute to
$j(x)$, and their effects can be similarly calculated).  
Contracting the down quark pair, and
using the algebraic relation of the SU(3) generators $t^a t^b = \frac{1}{N_{c}}
\delta_{ab} + \frac{1}{2}(f^{abc} + i d^{abc}) t^c $,  one then obtains  

\begin{equation} 
A_{O_{2}^{c}}= \frac{1}{N_{c}}\ \int dx \int \frac{dq}{(4\pi)^{4}} 
\frac{e^{-iqx}}{q^{2} -m_{d}^{2}} 
< \pi^{+}(p_{+}) \pi^{-}(p_{-}) | J^{\nu\mu}(x,q)J_{\nu\mu}^{G} | B(p) > 
\end{equation}
with 
\begin{eqnarray}
& & J^{\nu\mu} = \bar{d}(x)\gamma^{\nu}(q^{\rho}\gamma_{\rho} + m_{d}) 
\gamma^{\mu}
(1-\gamma_{5}) u(0), \nonumber \\ 
& & J^{G}_{\nu\mu} = \int_{0}^{1} d\alpha\ \alpha\ x^{\rho} 
\bar{u}(0) G_{\rho \nu}(\alpha x)
\gamma_{\mu} (1-\gamma_{5}) b(0) .
\end{eqnarray}
where we have neglected the terms given by two color-octet currents since their
nonzero contributions involve the next-to-leading order contributions of the
coupling constant $g_{s}$. 
 
  At this level, one can again use the factorization approach to evaluate the 
hadronic matrix element, and obtain

\begin{eqnarray}  
A_{O_{2}^{c}} & = & \frac{1}{N_{c}}\ \int dx \int \frac{dq}{(4\pi)^{4}}
\frac{e^{-iqx}}{q^{2} -m_{d}^{2}} \nonumber \\ 
& & \cdot < \pi^{-}(p_{-}) | J^{\nu\mu}(x,q) | 0 > 
< \pi^{+}(p_{+})| J_{\nu\mu}^{G} | B(p) > 
\end{eqnarray}
As the currents $J_{\mu\nu}(x,q)$ and $J^{G}_{\mu\nu}$ are nonlocal, their
hadronic matrix elements will be evaluated by so-called twist wave functions

\begin{eqnarray}
T^{\nu\mu}(p_{-},q,x) & = & 
< \pi^{-}(p_{-}) | J^{\nu\mu}(x,q) | 0 > =   if_{\pi} \int_{0}^{1} 
d\alpha\ e^{i\alpha x\cdot p_{-}} \nonumber \\
& & \cdot  [ (p_{-}^{\nu}q^{\mu} + 
p_{-}^{\mu}q^{\nu} -q\cdot p_{-} g^{\mu\nu} - i 
\varepsilon^{\nu\mu\rho\sigma}
q_{\rho}p_{-\sigma} )\ \phi_{A}(\alpha) \\
& & + m_{\pi}^{2} \frac{m_{d}}{m_{d} + m_{u}} ( g^{\nu\mu} \phi_{P}(\alpha)
 - i (p_{-}^{\mu}x^{\nu} - p_{-}^{\nu}x^{\mu}) \phi_{T}(\alpha) ) ] 
 \nonumber 
\end{eqnarray}
where $\phi_{A}$, $\phi_{P}$ and $\phi_{T}$ are pion twist wave functions
defined as follows\cite{CZ,BF}
\begin{eqnarray}
< \pi^-(p_{-}) | \bar{u}(x) i\gamma_{5} d(0) | 0 > & = &
\frac{f_{\pi}m_{\pi}^{2}}{m_{u}+m_{d}} \int_{0}^{1}d\alpha\ e^{i\alpha
x\cdot p_{-}} \phi_{P}(\alpha ) \\
 < \pi^-(p_{-}) | \bar{u}(x) \sigma_{\mu\nu} \gamma_{5} d(0) | 0 > & = &
\frac{-if_{\pi}m_{\pi}^{2}}{m_{u}+m_{d}} (p_{-\mu}x_{\nu} - p_{-\nu}x_{\mu})
\nonumber \\ 
& & \int_{0}^{1}d\alpha\ e^{i\alpha
x\cdot p_{-}} \phi_{T}(\alpha ) \\
< \pi^-(p_{-}) | \bar{u}(x) \gamma_{\mu}\gamma_{5} d(0) | 0 > & = & -i 
f_{\pi} p_{-\mu} \int_{0}^{1}d\alpha\ e^{i\alpha
x\cdot p_{-}} \phi_{A}(\alpha ) 
\end{eqnarray}
The twist wave functions $\phi_{P}$, $\phi_{T}$ and $\phi_{A}$ satisfy  the
normalization conditions
\begin{equation} 
\int_{0}^{1}d\alpha\ \phi_{A}(\alpha) =\int_{0}^{1}d\alpha\ 
 \phi_{P}(\alpha) =  \int_{0}^{1}d\alpha\ \phi_{T}(\alpha) =1
\end{equation}
These wave functions can be obtained from QCD sum rules. The results
depend on the renormalization group scale. For $\mu^{2} = m_{B}^{2} - m_{b}^{2}
= 5$ GeV$^{2}$, they are given by\cite{CZ}: $\phi_{A}(\alpha, \mu) = 
19.32 \alpha (1-\alpha)[(2\alpha - 1)^{2} + 0.11]$, $\phi_{P}(\alpha, \mu) 
= \frac{3}{4}[ 1.078 + 0.766(2\alpha - 1)^{2}]$, $\phi_{T}(\alpha, \mu) = 
\alpha (1- \alpha)$.

  To find the matrix element $< \pi^{+}(p_{+})| J_{\nu\mu}^{G} | B(p) >$, 
let us consider the correlator
\begin{eqnarray}
T_{\nu\mu}^{G}(p_{+},p,x) & = & i\int dy e^{-ipy}
 < \pi^{+}(p_{+})| J_{\nu\mu}^{G}\ \bar{b}(y)i\gamma_{5} d(y) |0 >  \\
& = & \frac{m_{B}^{2}f_{B}}{m_{b}} \frac{1}{m_{B}^{2} - p^{2}} 
 < \pi^{+}(p_{+})| J_{\nu\mu}^{G} | B(p) > + \cdots \nonumber
\end{eqnarray}
where $<B(p)|\bar{b}(y)i\gamma_{5}d(y) |0> = (m_{B}^{2}f_{B}/m_{b})e^{-ipy}$.
The ellipses denote the excited and continuum states which are usually taken
into account by the dispersion integral with the effective threshold. 
The single soft gluon contribution is given by 
\begin{eqnarray}
T_{\nu\mu}^{G}(p_{+},p,x) & = & i\int dy \int dk 
\frac{e^{-i(p-k)y}}{(4\pi)^{4}(m_{b}^{2}-k^{2})} 
\int_{0}^{1} d\alpha_0\ \alpha_0\ x^{\rho}  \\
 & & \cdot < \pi^{+}(p_{+})|\bar{u}(0) G_{\rho \nu}(\alpha_0 x)
\gamma_{\mu}\gamma_{5}(k^{\sigma}\gamma_{\sigma} + m_{b})\gamma_{5} d(y) | 0> 
\nonumber 
\end{eqnarray}
which can be evaluated by introducing the pion wave functions with higher twist
\begin{eqnarray}
T_{\nu\mu}^{G}(p_{+},p,x) & & = i\int dy \int dk
\frac{e^{-i(p-k)y}}{(4\pi)^{4}(m_{b}^{2}-k^{2})}\int_{0}^{1}d\alpha_0 \alpha_0
\{  \nonumber \\
& & f_{\pi} \delta_{3\pi} \int D\alpha_{i} \phi_{3\pi}(\alpha_{i})
e^{ip_{+}(\alpha_{2}y + \alpha_{3}\alpha_0 x)} \nonumber \\
& & \cdot [ (k_{\nu} x\cdot p_{+}- p_{+\nu}x\cdot k)
p_{+\mu} - (g_{\nu\mu} x\cdot p_{+} - p_{+\nu}x_{\mu}) k\cdot p_{+} ]
 \\
& & + f_{\pi}\delta_{4\pi}^{2}m_{b} \int D\alpha_{i}   
e^{ip_{+}(\alpha_{2}y + \alpha_{3}\alpha x)} 
[ \phi_{\perp}(\alpha_{i}) (g_{\nu\mu} x\cdot p_{+} - p_{+\nu}x_{\mu})
 \nonumber \\
& & + (\phi_{\parallel}(\alpha_{i}) + \phi_{\perp}(\alpha_{i}) )
(p_{+\nu} \frac{y\cdot x}{y\cdot p_{+}} -
\frac{x\cdot p_{+}}{y\cdot p_{+}}y_{\nu})
(p_{+\mu} - \alpha_{2}k_{\mu}\frac{m_{\pi}^{2}}{m_{d}m_{b}})]\} \nonumber 
\end{eqnarray}
where $\phi_{3\pi}$ and $\phi_{\parallel}$, $\phi_{\perp}$ are the pion wave
functions of twist 3 and 4, respectively.  They are defined as \cite{CZ,BF}
\begin{eqnarray}
& & < \pi^{+}(q)|  \bar{u}(0) G_{\rho \nu}\sigma_{\mu\sigma} 
\gamma_{5} d(y) | 0> = -i\int D\alpha_{i}
\phi_{3\pi}(\alpha_{i}) e^{iq(\alpha_{2}y + \alpha_{3}\alpha_0 x)} \nonumber \\
& & \cdot f_{\pi}\delta_{3\pi} 
 [ q_{\mu} (q_{\rho}g_{\nu\sigma} - q_{\nu}g_{\rho\sigma}) -
q_{\sigma}(q_{\rho}g_{\nu\mu} - q_{\nu}g_{\rho\mu})]; \\
& &  < \pi^{+}(q)| \bar{u}(0) G_{\rho \nu}\gamma_{\mu}
\gamma_{5} d(y) | 0> = \int D\alpha_{i}\phi_{\perp}(\alpha_{i}) 
e^{iq(\alpha_{2}y + \alpha_{3}\alpha_0 x)} \nonumber \\
& & \cdot f_{\pi}\delta_{4\pi}^{2}  
\cdot [ q_{\nu} (g_{\rho\mu} - \frac{y_{\rho}q_{\mu}}{y\cdot q}) -
q_{\rho}(g_{\nu\mu} - \frac{y_{\nu}q_{\mu}}{y\cdot q})] 
 \nonumber \\
& & +  f_{\pi}\delta_{4\pi}^{2} q_{\mu} \frac{q_{\rho}y_{\nu}
-q_{\nu}y_{\rho}}{y\cdot q} \int D\alpha_{i}\phi_{\parallel}(\alpha_{i}) 
e^{iq(\alpha_{2}y + \alpha_{3}\alpha_0 x)}
\end{eqnarray}
where $D\alpha_{i} = d\alpha_1 d\alpha_2 d\alpha_3 \delta(\alpha_1 + \alpha_2 
+ \alpha_3 -1)$. The constants $\delta_{3\pi}$ and $\delta_{4\pi}$ are
renormalization scale dependent and have the values\cite{BF} 
$\delta_{3\pi}\simeq 0.027$GeV and 
$\delta_{4\pi}\simeq 0.45$ GeV at $\mu \simeq 1$ GeV. The results for 
$\phi_{3\pi}$, $\phi_{\perp}$ and $\phi_{\parallel}$  are given by 
$\phi_{3\pi}(\alpha_{i}) = 360 \alpha_1 \alpha_2 \alpha_3^{2}  
[ 1 + \omega_{1,0}\frac{1}{2}(7\alpha_{3} -3)+ 
\omega_{2,0}(2-4\alpha_1\alpha_2 -8\alpha_3 + 8\alpha_{3}^{2})+ 
\omega_{1,1} (3\alpha_1 \alpha_2 - 2\alpha_3 + 3\alpha_3^{2}) + \cdots ]$,
$\phi_{\perp}(\alpha_i)=30 (\alpha_1 - \alpha_{2})
\alpha_{3}^{2} [\frac{1}{3} + 2\varepsilon (1 - 2\alpha_3) ]$,
$\phi_{\parallel}(\alpha_i)=120 \varepsilon (\alpha_1 - \alpha_{2})
\alpha_1 \alpha_2 \alpha_3\ $.
where $\omega_{1,0} = -2.88$, $\omega_{2,0} = 10.5$, $\omega_{1,1} = 0$ and 
$\varepsilon \simeq 0.5$ at $\mu \simeq 1$ GeV. 
  
 The integral of eq. (18) for the variables $y$ and $k$ can be performed
 and is given by
\begin{eqnarray}
T_{\nu\mu}^{G}(p_{+},p,x) & & = i \int D\alpha_{i}
\frac{f_{\pi}}{m_{b}^{2}-(p-\alpha_{2}p_{+})^{2})} \int_{0}^{1}d\alpha_0
\alpha_0 e^{i\alpha_0 \alpha_3 p_{+}\cdot x} \nonumber \\
 & & \{\delta_{3\pi}\phi_{3\pi}(\alpha_{i})  
(p_{\nu} x\cdot p_{+} -p_{+\nu}x\cdot p) p_{+\mu} - 
(g_{\nu\mu} x\cdot p_{+} - p_{+\nu} x_{\mu}) 
\nonumber \\
& & \cdot \left( (p\cdot p_{+} -\alpha_{2} p_{+}^{2})
\delta_{3\pi} \phi_{3\pi}(\alpha_{i}) -
\delta_{4\pi}^{2} m_{b}\phi_{\perp}(\alpha_{i}) \right) \\ 
& & + \delta_{4\pi}^{2}m_{b} 
\left(\phi_{\perp}(\alpha_{i}) +  \phi_{\parallel}(\alpha_{i})\right)
\int_{0}^{\infty} d\beta \frac{m_{b}^{2} -(p-\alpha_{2}p_{+})^{2}}{
m_{b}^{2} -(p-(\alpha_{2}+\beta)p_{+})^{2}}  \nonumber \\ 
& & \cdot [  \frac{m_{\pi}^{2}}{m_{d}m_{b}}
(g_{\nu\mu} x\cdot p_{+} - p_{+\nu} x_{\mu}) + \frac{2}{
m_{b}^{2} -(p-(\alpha_{2}+\beta)p_{+})^{2}} \nonumber \\
& & \cdot (p_{\nu} x\cdot p_{+} - p_{+\nu}x\cdot p)\left( p_{+\mu} + 
\frac{m_{\pi}^{2}}{m_{d}m_{b}}\alpha_{2}((\alpha_{2} + 
\beta) p_{+\mu} + p_{\mu})\right) ] \}
\nonumber
\end{eqnarray}
Here we have used the relation $1/p_{+}\cdot y = -i\int_{0}^{\infty} d\beta 
e^{i\beta p_{+}\cdot y}$. With the above results, we further consider 
\begin{eqnarray}
& & \frac{1}{N_{c}}\ \int dx \int \frac{dq}{(4\pi)^{4}}
\frac{e^{-iqx}}{q^{2} -m_{d}^{2}}\
T^{\nu\mu}(p_{-},q,x)T^{G}_{\nu\mu}(p_{+},p,x) \nonumber \\
& & = \frac{m_{B}^{2}f_{B}}{m_{b}}\frac{1}{m_{B}^{2}-p^{2}} A_{O_{2}^{c}} 
+ \int_{s_{0}}^{\infty} \frac{\rho (m_{\pi}^{2}, s) ds }{s - p^{2}}
\end{eqnarray} 
where the dispersion integral with the effective threshold $s_{0}$ assembles
the excited and continuum states. 
By substituding eqs.(10) and (24) to the left hand side of the above equation,  
it is not difficult, though tedious, to integrate out the variables 
$x$ and $q$. The result is found to be 

\begin{eqnarray}
& & \frac{1}{N_{c}}\ \int dx \int \frac{dq}{(4\pi)^{4}}
\frac{e^{-iqx}}{q^{2} -m_{d}^{2}}\
 T^{\nu\mu}(p_{-},q,x)T^{G}_{\nu\mu}(p_{+},p,x) \nonumber \\
& & = i \frac{1}{N_{c}} f_{\pi}^{2} \int_{0}^{1}d\alpha \int_{0}^{1}d\alpha_0\ 
\alpha_0 \int D\alpha_{i} \frac{\phi_{A}(\alpha)}{m_{b}^{2} - (p-\alpha_2
p_{+})^{2}} \{  \nonumber \\
& & \frac{p^{2}}{2[(\alpha - \alpha_0 \alpha_3)^{2} 
m_{\pi}^{2} + \alpha \alpha_0
\alpha_3 p^{2} - m_{d}^{2}]^{2}} [ \alpha (\alpha_0 \alpha_3 p^{2} 
+ 2(\alpha -\alpha_0\alpha_3)m_{\pi}^{2}) \nonumber \\
& & \cdot ( \delta_{3\pi} \phi_{3\pi}(\alpha_{i}) p^{2} 
- \delta_{4\pi}^{2} m_{b}
\phi_{\perp}(\alpha_i) + \delta_{4\pi}^{2} m_{b}(\phi_{\perp}(\alpha_i) 
+ \phi_{\parallel}(\alpha_i) ) \nonumber \\
& & \cdot \int_{0}^{\infty} d\beta \frac{m_{b}^{2} - 
(p-\alpha_2 p_{+})^{2}}{m_{b}^{2} - (p-(\alpha_2 + \beta) p_{+})^{2}} 
 ( \frac{p^{2}}{m_{b}^{2} - (p-(\alpha_2 + \beta) p_{+})^{2}} \\
& & \cdot ( 1 + \frac{m_{\pi}^{2}}{m_{d}m_{b}} \alpha_2 (1 
+ \alpha_2 + \beta) )  - \frac{m_{\pi}^{2}}{m_{d}m_{b}} ) )
  \nonumber \\
& &  +\left( (\alpha - \alpha_0 \alpha_3)^{2} m_{\pi}^{2} + 
\alpha \alpha_0 \alpha_3 p^{2} \right) 
(  \delta_{3\pi} \phi_{3\pi}(\alpha_{i}) p^{2} 
- 2\delta_{4\pi}^{2} m_{b}\phi_{\perp}(\alpha_i) \nonumber \\
& & -2 \delta_{4\pi}^{2} m_{b}(\phi_{\perp}(\alpha_i) 
+ \phi_{\parallel}(\alpha_i) )
\int_{0}^{\infty} d\beta \frac{m_{b}^{2} - 
(p-\alpha_2 p_{+})^{2}}{m_{b}^{2} - (p-(\alpha_2 + \beta) p_{+})^{2}} 
\frac{m_{\pi}^{2}}{m_{d}m_{b}} ) ] \nonumber \\
& & + \frac{p^{2}}{(\alpha - \alpha_0 \alpha_3)^{2} 
m_{\pi}^{2} + \alpha \alpha_0\alpha_3 p^{2} - m_{d}^{2}}[2\delta_{3\pi} 
\phi_{3\pi}(\alpha_{i}) p^{2} 
- 3\delta_{4\pi}^{2} m_{b}\phi_{\perp}(\alpha_i)  \nonumber \\
& & + \delta_{4\pi}^{2} m_{b}(\phi_{\perp}(\alpha_i) 
+ \phi_{\parallel}(\alpha_i))\int_{0}^{\infty} d\beta \frac{m_{b}^{2} 
- (p-\alpha_2 p_{+})^{2}}{m_{b}^{2} - 
(p-(\alpha_2 + \beta) p_{+})^{2}} \nonumber \\ 
& & \cdot \left( ( 1 + \frac{m_{\pi}^{2}}{m_{d}m_{b}} \alpha_2 ( 1+ 
\alpha_2 \beta) )
 \frac{p^{2}}{m_{b}^{2} - (p - (\alpha_2 + \beta)p_{+})^{2}} - 3
\frac{m_{\pi}^{2}}{m_{d}m_{b}} \right) ] \}  \nonumber 
\end{eqnarray}
where we have neglected the terms of order $m_{\pi}^{2}/m_{B}^{2}$. 
The remaining integral may be carried out numerically (in fact, the 
integral for the variable $\beta$ can be explicitly integrated out). 
Before doing so, one may use the QCD sum rule approach\cite{SVZ} by
applying the Borel transformation in the variable $p^{2}$,  i.e.,
\begin{equation}
\hat{L}{}_{M^{2}} f(Q^{2}) 
= \lim_{Q^{2}, n \rightarrow \infty, (Q^{2}/n)=M^{2}} 
\frac{(Q^{2})^{(n+1)}}{n!} \left( - \frac{d}{dQ^{2}}\right )^{n} f(Q^{2})
\equiv f(M^{2})
\end{equation}
to the left and right hand side of eq.(25) with use of the more explicit 
expression eq.(26) for the left hand side. 
$M$ is the so-called Borel parameter. It can be seen that the Borel
operator transforms the series in $Q^{-2}$ into a series in $M^{-2}$: 

\begin{equation}
\hat{L}_{M} \left(\frac{1}{Q^{2}}\right)^{n} = \frac{1}{(n-1)!} 
\left(\frac{1}{M^{2}}\right)^{n} 
\end{equation}
 
  From duality considerations, the dispersion integral on the right hand side 
of eq.(25) is cancelled against the corresponding part of the dispersion 
integral of the QCD result on the left hand side.
One then reads off the needed amplitude $A_{O_{2}^{c}}$. The procedure is quite 
standard following the usual QCD sum rule calculations\cite{SVZ}. 
The utility of the result is that it depends on twist wave functions that can be
calculated from two point Greens' functions. We have used $B^0 \rightarrow
\pi^+ \pi^-$ as an example but the method applies equally well to other heavy
to light channels. A detailed  application will be discussed elsewhere.

  In conclusion, we have shown that the nonfactorizable contributions to the
hadronic matrix elements of the nonleptonic two body decays can be calculated 
directly from the QCD theory. It has been seen that the  
leading nonzero contributions 
from the single soft gluon emission can actually be factorized again 
in terms of the higher twist wave functions of the mesons. 
It is expected that the approach presented here will be helpful in the 
understanding hadronic decays of the
heavy mesons, especially for the heavy to light meson decays.
It shall provide a complement to the approaches presented in ref.\cite{BD} for 
the calculations of the $B\rightarrow D\pi$ and $B\rightarrow J/\psi K$ 
decays in which one of the final hadrons is heavy.

%\begin{references}

\end{document}